\documentclass[aps,twocolumn,prb,superscriptaddress]{revtex4}
\usepackage{epsfig}
\usepackage{amssymb}
\usepackage{times}
\begin{document} 
  
\title{Polarization Enhancement in Short Period Superlattices via Interfacial Intermixing} 

\author{Valentino R. Cooper}
\affiliation{Department of Physics and Astronomy, Rutgers University,
136 Frelinghuysen Rd, Piscataway, New Jersey 08854-8019, USA}
\email{vcooper@physics.rutgers.edu} 

\author{Karen Johnston}
\affiliation{Laboratory of Physics, Helsinki University of Technology,
P.O. Box 1100, 02015, Finland}

\author{Karin M. Rabe} 
\affiliation{Department of Physics and Astronomy, Rutgers University,
136 Frelinghuysen Rd, Piscataway, New Jersey 08854-8019, USA}

\date{\today}

\begin{abstract}

The effect of intermixing at the interface of short period
PbTiO$_3$/SrTiO$_3$ superlattices is studied using first-principles
density functional theory.  The results indicate that interfacial
intermixing significantly enhances the polarization within the
superlattice.  This enhancement is directly related to the
off-centering of Pb and Sr cations and can be explained through a
discussion of interacting dipoles.  This picture should be general for
a wide range of multicomponent superlattices and may have important
consequences for the design of ferroelectric devices.

\end{abstract}

\maketitle 

Ferroelectric superlattices are a promising paradigm for creating
novel materials for device applications.  In ideal superlattices with
perfectly flat, compositionally abrupt interfaces, first-principles
calculations and experimental studies have shown how factors such as
strain due to lattice mismatches~\cite{Tabata94p1970, Christen96p1488,
Jiang99p2851, LeMarrec00pR6447, Nakagawara00p3257, Sepliarsky01p4509,
Jiang03p1180, Neaton03p1586, Rios03pL305, Bungaro04p184101,
Dawber05p177601, Johnston05p100103, Lee05p395, Nakhmanson05p102906},
charge compensation~\cite{Dawber03pL393, Junquera03p506, Ahn04p488,
Kolpak06p054112} and bonding at the interface~\cite{Fong06p127601} can
be controlled to enhance the ferroelectric properties of the
superlattice. Work by Dawber and coworkers show that the behavior of
very short period superlattices may deviate from that extrapolated
from longer periods~\cite{Dawber05p177601}.  While, these effects may
be the result of a number of different factors, it seems likely that
changes at the interface dominate at these length scales.  Recent high
resolution COBRA studies on SrTiO$_3$ (STO) supported PbTiO$_3$ (PTO)
thin films further suggest that cation intermixing may be present at
the interface of such superlattices~\cite{Fong05p144112}.

Knowledge of how interfacial cation intermixing influences the
aggregate properties of a superlattice can be useful in tuning its
properties.  As the superlattice period decreases, these effects
become more important and, depending on their strength, may dominate.
In this paper, we present a first principles study on the effect of
interfacial cation intermixing on the polarization of short period
PTO/STO superlattices.  Our results demonstrate that intermixing
significantly enhances the polarization and that this enhancement is
directly linked to larger Pb displacements in the intermixed layers.

\begin{figure}
\includegraphics[height=1.75in]{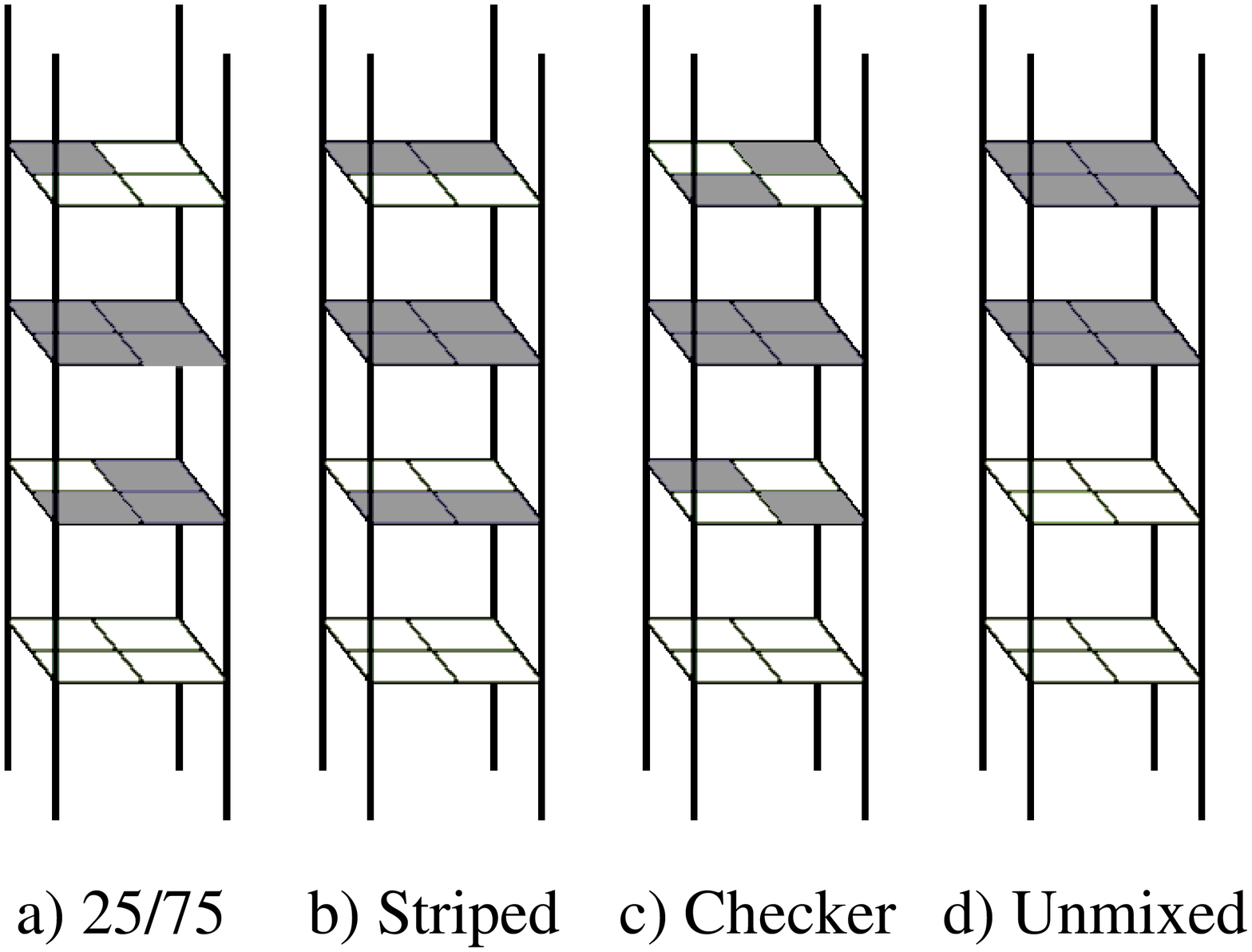}
\caption{\label{mixed_layers} Representative 2 PTO / 2 STO structures
with intermixed interfaces.  Grey squares are PTO unit cells and white
squares are STO. Intermixed layers have a total composition of 50$\%$
PTO / 50$\%$ STO.  The two intermixed layers account for 1 PTO and 1
STO layer.}
\end{figure}

We use density functional theory (DFT) to examine the effects of
interlayer cation mixing on the polarization within PTO/STO
superlattices.  We compare superlattices with sharp interfaces with
three different compositionally equivalent intermixed systems of m=1-3
PTO layers and n=1-3 STO layers
(figure~\ref{mixed_layers}). Checkered, striped and 25\%/75\%
intermixed layers were studied. Intermixed layers were placed at each
of the mixed PTO/STO interfaces; giving a total composition of 50$\%$
PTO / 50$\%$ STO.  The two intermixed layers account for one PTO and
one STO layer. All calculations were performed using projector
augmented wave (PAW) potentials~\cite{Blochl90p5414, Kresse99p1758}
with the Vienna Ab initio Simulation Package (VASP
v4.6.26)~\cite{Kresse96p11169}, with the local density approximation
for the exchange correlation functional. A 600 eV (22 Ha) cutoff and a
4$\times$4$\times$4 k point mesh were used. STO was found to be cubic
with a lattice constant of 3.863~\AA\ (experiment:
3.905~\AA~\cite{Hellwege1981}). The tetragonal PTO lattice constants
were computed as $a$=3.867~\AA\ and $c$=4.033~\AA\ (experiment:
$a$=3.904~\AA\ and $c$=4.152~\AA~\cite{Hellwege1981}). This agreement
is typical of LDA calculations for ferroelectric perovskites. In all
calculations the in-plane lattice constant $a$ of the superlattice was
constrained to that of the STO substrate (3.863~\AA) while the $c$
lattice vectors were optimized within the $P4mm$ space group. All
ionic coordinates were fully relaxed until the Hellman-Feynman forces
on the ions were less than 5 meV/\AA. For bulk PTO constrained to the
STO in-plane lattice constant of 3.863~\AA\ the $c$ axis lattice
parameter was 4.039~\AA\ and the polarization was 0.79 C/m$^2$.
Polarizations were computed using the Berry phase
method~\cite{King-Smith93p1651}.

\begin{figure}[t]
\includegraphics[height=1.75in]{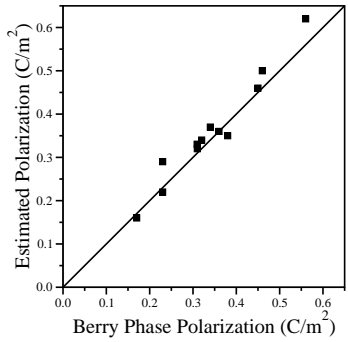}
\caption{\label{BerryPol} A comparison between the polarization
calculated using the Berry phase method and bulk $Z^*$s and cation
off-centering.  The solid squares represent individual data points
taken from the subset of calculations for PTO/STO superlattices
with and without intermixing.  }
\end{figure}

Figure~\ref{BerryPol} shows a comparison between the Berry phase
polarization and the polarization calculated using bulk Born effective
charges (Z$^*$) and cation off-centering for all the intermixed and
unmixed superlattices studied. Cation displacements were computed from
the center of their respective oxygen cages (i.e. 6 nearest oxygen
neighbors for Ti and 12 for Pb and Sr). The bulk $Z^*$s were:
Z$^*_{\rm Pb,Sr}$ = 2.7 and Z$^*_{\rm Ti}$ = 4.6.  The remarkable
agreement between both methods suggests that the bulk $Z^*$s remain
relatively unchanged in all superlattices.  This analysis allows us to
relate changes in the total polarization to the off-center
displacements of the individual cations.

\begin{figure}[b]
\includegraphics[height=1.75in]{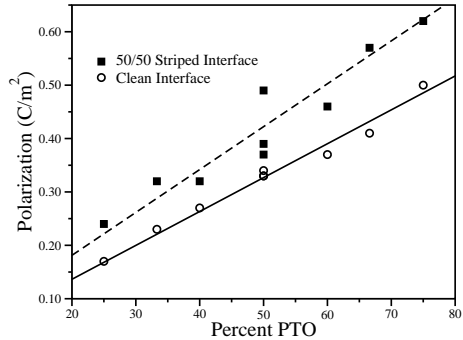}
\caption{\label{Polarization} Polarization as a function of PTO
concentration for m$_{\rm PTO}$/n$_{\rm STO}$ short period
superlattices.  Solid squares ($\blacksquare$) represent striped
interfaces and open circles ($\circ$) are for sharp
interfaces. The solid (---) and dashed lines(- -) are fits through the
data.}
\end{figure}

\begin{figure}
\includegraphics[height=1.70in]{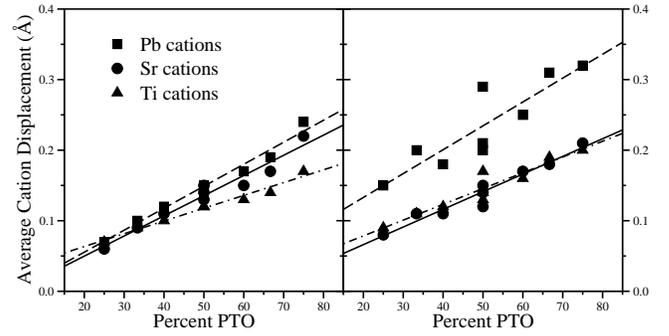}
\caption{\label{cation_disp} Average cation displacements as a function
of PTO concentration for the unmixed (left) and striped (right)
interfaces.  Squares ($\blacksquare$) represent Pb data, circles
($\bullet$) are for Sr and triangles ($\blacktriangle$) are Ti.  Solid
and dashed lines are fits to the data. Pb off-center displacements are
greatly enhanced in superlattices with intermixing at the interface.}
\end{figure}

In figure~\ref{Polarization} we see that the striped interface shows a
marked increase in the total polarization as a function of percent PTO
for nearly all short-period superlattices studied.  While the striped
interface is certainly the most dramatic example of polarization
enhancement in an intermixed system, both the 25/75 and checkered
interfaces display significant increases in their polarization
relative to the sharp interface, suggesting that this trend is the
consequence of interfacial cation intermixing.  This polarization
enhancement can be understood by considering the magnitude of the
average cation displacements in both intermixed and sharp
superlattices.  Figure~\ref{cation_disp} depicts the average cation
off-center displacements for Pb, Sr and Ti ions as a function of PTO
concentration for superlattices with striped (right) and unmixed
interfaces (left).  These results indicate little change in the
average Sr and Ti off-center displacements and clearly demonstrate
that the changes in the polarization within the intermixed
superlattices are due to changes in the off-centering of Pb cations.
Once again, this trend is common to all intermixed configurations
studied.

\begin{figure}[b]
\includegraphics[height=1.75in]{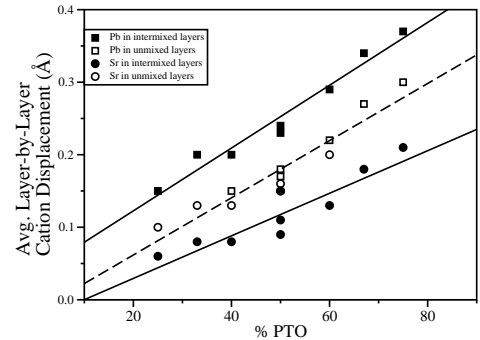}
\caption{\label{PbandSr_intermix} 
Average layer-by-layer cation displacements as a function of PTO
concentration in intermixed layers and unmixed layers for PTO/STO
superlattices with striped interfaces.  Solid squares ($\blacksquare$)
and circles ($\bullet$) represent intermixed Pb and Sr layers. While,
open squares ($\Box$) and circles ($\circ$) are for unmixed Pb
and Sr layers.  Solid (---) and dashed lines (- -) are guides for
intermixed and unmixed data, respectively.}
\end{figure}

To further elucidate the changes induced by interfacial cation
intermixing we compared the changes in the average layer-by-layer
cation off-centering for Pb and Sr cations in striped interfaces to
those in pure layers (Figure~\ref{PbandSr_intermix}).  Here we see
that the magnitude of the average layer-by-layer Pb off-center
displacements in unmixed Pb layers is essentially equal to that of Sr
cations in unmixed Sr layers. In addition, the average layer-by-layer
Sr cation off-center displacements decrease when comparing an unmixed
layer to an intermixed layer.  Conversely, Pb cations off-center more
in the intermixed layers.

Now we turn to the discussion of how the differences between unmixed
and intermixed superlattices arise.  First, we recall the present
understanding of polarization in unmixed superlattices.  PTO is a
well-studied ferroelectric material with a double well potential which
results in a spontaneous polarization in zero electric field.  STO, on
the other hand, is an incipient ferroelectric with a very shallow
potential energy well centered around the paraelectric state. Previous
first principles superlattice studies have shown that in ideal
ferroelectric-paraelectric superlattices with compositionally sharp
interfaces the ferroelectric layers, e.g. PTO, with polarization along
the stacking direction, induce a polarization within the paraelectric
layers.  Furthermore, the magnitude of the layer-by-layer polarization
throughout the superlattice was nearly
constant~\cite{Neaton03p1586, Nakhmanson05p102906, Wu06p107602}; a
direct consequence of the minimization of the electrostatic energy
associated with the buildup of polarization charge at the
interfaces~\cite{Neaton03p1586}. The polarization of the superlattice
is then intermediate between zero and the full spontaneous
polarization of the ferroelectric layer material, increasing with
increasing layer fraction of the ferroelectric phase. Our results for
unmixed short-period PTO/STO superlattices are fully consistent with
these previous results. In particular, the STO and PTO layers have
nearly the same polarization, less than the full bulk spontaneous
polarization of PTO, which increases with the compositional fraction
of PTO. Since the $Z^*$s of Sr and Pb in STO and PTO are very similar,
the Sr and Pb displacements are correspondingly nearly equal.

\begin{figure} \includegraphics[height=2.50in]{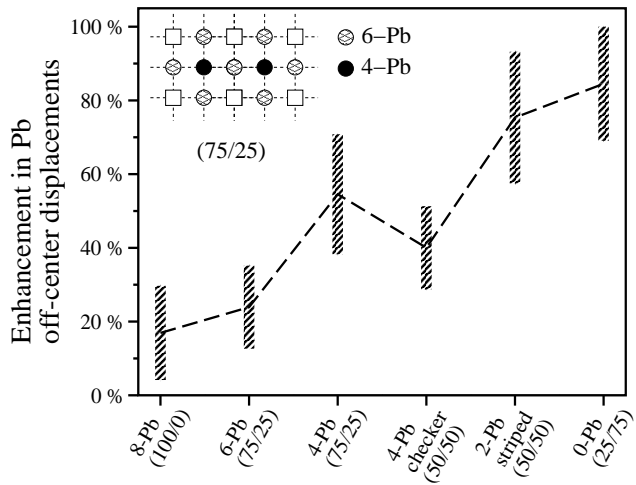}
\caption{\label{nearPb} The percent enhancement in Pb cation
off-centering as a function of number of Pb nearest neighbors.  Shaded
bars represent the average and standard deviations for superlattices
with various periods and compositions.  The values in parentheses
represent the percent PTO/STO in the layer. The inset depicts the
arrangement of the two types of Pb cations in the 75 PTO/25 PTO layer.
Empty squares ($\square$) are for Sr cations, filled circles
($\bullet$) represent Pb cations with four nearest neighbors, and
hashed circles are Pb cations with 6 in-plane neighbors (2 nearest
neighbors and 4 second nearest neighbors). This plot was generated
using data from all three intermixed superlattices as well as unmixed
superlattices.}
\end{figure}

Next, to understand the effects of intermixing we assign to each PTO
or STO unit cell a single dipole, corresponding to the local polar
distortion. Each layer is to be regarded as a square planar array of
dipoles.  In an unmixed layer, where the dipoles are all PTO or all
STO, the magnitudes of all dipoles in that layer are the same.  On the
other hand, in an intermixed layer where the potential energy wells
governing the dipole moments of the two types of unit cells are
significantly different, we expect the magnitudes of the dipoles to be
different. Without changing the net polarization of the layer, the PTO
unit cell dipoles can be larger, and the STO unit cell dipoles
smaller, reducing the energy cost for maintaining the polarization. As
we discuss further below, depending on the details of the potential
wells and the Pb/Sr arrangement in the layer, this gain in energy may
drive an increase in the overall polarization of the layer, and
consequently of the superlattice.

A more precise understanding of this argument can be obtained by
examining the Pb off-center displacements as a function of the number
of nearest in-plane Pb neighbors.  To isolate the intermixing effect,
we report these values relative to the average Pb displacements in
compositionally identical unmixed superlattices with the same period.
As seen in Figure~\ref{nearPb}, the observed enhancement depends on
the composition of the intermixed layer and the number of Pb in-plane
neighbors in the layer.  First, there is a rough negative correlation
between the Pb fraction in the intermixed layer and the enhancement of
the Pb off-centering. Second, at a given Pb fraction there is a
negative correlation between the number of Pb in-plane neighbors and
enhancement of the Pb off-center displacements.

These trends can be understood by considering the dipole-dipole
repulsions within the intermixed layer. Each pair of aligned Pb
dipoles costs electrostatic energy, which increases as the spacing
between the dipoles decreases or as the magnitude of the dipoles
increases. A smaller Pb fraction in the layer decreases the relative
importance of this contribution, favoring an increase in the magnitude
of the Pb dipole. At a fixed fraction, the size of this energy will
depend on the arrangement of the Pb dipoles in the layer, with a
smaller number of close (first and second) Pb-Pb neighbors decreasing
the relative importance of this contribution and similarly favoring an
increase in the magnitude of the Pb dipole. This final point is
exemplified by considering the 75/25 mixed layer (see
Figure~\ref{nearPb} inset).  Here we see that there are two distinct
types of Pb cations, the 4-Pb and the 6-Pb cation.
Figure~\ref{nearPb} shows that in this case the 4-Pb cation has a much
larger enhancement, consistent with the decreases in dipole-dipole
repulsions with decreasing first and second Pb nearest in-plane
neighbors.  Similarly, Pb dipoles in striped (2-Pb nearest neighbors)
layers are enhanced relative to those in the compositionally
equivalent checkered (4-Pb neighbors) layers. In all cases, larger Pb
dipoles in the intermixed layers increase the average polarization of
that layer, which, through the same electrostatic arguments presented
for the sharp interface superlattices, favors an enhancement of the
total polarization.  From this it can be understood that the striped
interface superlattices, in which the Pb cations have fewer Pb
neighbors than the checkered and the combined 25/75-75/25
superlattices, should exhibit more enhanced polarization, as observed
in the first principles calculations.

\begin{table}
\begin{tabular}{ccccc}
\hline
\hline
& \multicolumn{3}{c}{$\Delta E_{\rm mixing}$ (meV/cell)}\\
\hline
$m_{\rm PTO}$/$n_{\rm STO}$ & 25/75 & Striped & Checker & Sharp\\
\hline 
\hline
1/1 &  33 & 27 & 32 & 45\\
1/2 & -6 & -7 & -5 & 5 \\
1/3 & 1 & 1 & 1 & 2\\
2/1 & -35 & -41 & -35 & -18\\
2/2 & -16 & -17 & -16 & -5\\
2/3 & -6 & -8 & -6 & -5\\
3/1 & -46 & -52 & -46 & -29\\
3/2 & -23 & -26 & -23 & -20\\
3/3 & -14 & -15 & -14 & -11\\
\hline
\hline
\end{tabular}
\caption{$\Delta E_{\rm mixing}$ relative to bulk for interfacial
intermixed and unmixed superlattices. In all cases intermixing is
preferred.}
\label{delta_mix}
\end{table}

Finally we consider whether interfacial cation intermixing is
energetically favorable.  Table~\ref{delta_mix} shows the $\Delta
E_{\rm mixing}$ relative to bulk PTO and STO for the three interfacial
intermixed cells and the sharp interface.  In all cases, we see that
intermixing is preferred over sharp interfaces.

In conclusion, our results show that intermixed PTO/STO interfaces
enhance the polarization of a superlattice relative to sharp
interfaces.  This enhanced polarization is strongly linked to larger
Pb displacements in the intermixed layers and can be rationalized
through a consideration of the dipole-dipole interactions.  Since this
model is based solely on the interaction of dipoles constrained by
potential energy wells, these concepts should be general to all
multicomponent superlattices in which there is a significant
difference in the ferroelectric and/or dielectric properties of the
substituent layers.  This study, combined with knowledge from previous
studies on strain and internal electric field effects in
superlattices, suggests a new mechanism for tailoring short period
superlattices for specific applications.

\end{document}